\documentclass[]{jfm}

\usepackage{graphicx}
\usepackage{newtxtext}
\usepackage{newtxmath}
\usepackage{natbib}
\usepackage{hyperref}
\usepackage{bm}

\hypersetup{
    colorlinks = true,
    urlcolor   = blue,
    citecolor  = black,
}

\newcommand{\RomanNumeralCaps}[1]
\linenumbers

\newcommand{\We}{\mathrm{We}}
\newcommand{\Bo}{\mathrm{Bo}}
\newcommand{\De}{\mathrm{De}}

\newcommand\norm[1]{\left\lVert#1\right\rVert}

\title{Wavemaker and endogeneity of gravitationally stretched weakly viscoelastic jets}

\author{Daniel Moreno-Boza
  \corresp{\email{damoreno@eng.uc3m.es}}
  \aff{1}
}

\affiliation{\aff{1}Grupo de Mecánica de Fluidos, Departamento de Ingeniería Térmica y Fluidos, University Carlos III, Madrid.
}

\begin{document}
\maketitle

\begin{abstract}
Highly stretched capillary jets produced by gravity are central to drop generation, micro-thread formation, and extensional-rheometry concepts. 
For Newtonian fluids, the transition from steady jetting to self-excited oscillations in a gravitationally stretched jet is predicted accurately by one-dimensional slender-jet equations that retain the exact interfacial curvature and admit a global eigenvalue analysis \citep{RubioRubio2013}. 
Separately, weakly viscoelastic jets governed by Oldroyd--B/Giesekus constitutive laws exhibit elastocapillary regimes and beads-on-a-string dynamics that are well captured by one-dimensional free-surface models \citep{Ardekani2010}. 
Here we formulate a unified one-dimensional model for gravitationally stretched viscoelastic jets, combining full-curvature capillarity with a Giesekus stress closure, and we analyse its global linear stability on spatially developing base states. 
We first benchmark the Newtonian limit, reproducing marginal spectra and base-flow profiles, and then quantify how elasticity shifts the critical jetting--dripping boundary by tracking the leading global Hopf eigenpair across the rheological parametric space. 
For experimentally relevant moderate elasticity, characterised by order-unity Deborah numbers, polymeric tension modifies both the critical Weber number and the selected oscillation frequency, and endogeneity decompositions reveal that marginality results from a balance between capillary/kinematic contributions and an additional elastic-stress feedback pathway. 
To interpret and predict the onset mechanism, we compute wavemakers and receptivity/structural-sensitivity fields from direct--adjoint eigenfunctions, showing that viscoelasticity broadens the sensitivity region downstream while the adjoint remains strongly localized near the inlet, thereby identifying the near-nozzle region as the dominant receptive location. 
\end{abstract}


{\bf MSC Codes }  {\it(Optional)} Please enter your MSC Codes here

\section{Introduction and motivation}

The controlled generation of micron- and sub-micron-sized threads, films and droplets is central to a wide range of technologies in medicine, pharmaceutics, chemical engineering and materials science \citep[e.g.][]{basaran2002small,stone2004engineering,barrero2007micro,utada2007b,EggersVillermaux2008}. 
A recurring strategy is to produce a liquid filament whose diameter is substantially smaller than that of the injector, thereby avoiding the fabrication and operation of micro-scale solid geometries. 
This idea underpins classical fibre-spinning configurations \citep{pearson1969spinning,denn1980continuous}, electrospinning devices where electrical stresses stretch the jet \citep{doshi1995electrospinning,loscertales2002micro,marin2007simple}, and flow-focusing/co-flow arrangements that generate thin inner jets and micro-drops through capillary breakup \citep{ganan1998generation,ganan2001perfectly,anna2003formation,suryo2006tip,utada2007b,marin2009generation,castro2012slender}. 
In this landscape, gravitational stretching provides a particularly robust and conceptually simple route: a millimetric nozzle can generate a long, slender thread that thins downstream under the combined action of gravity and continuity, producing droplets or micro-threads whose characteristic scale can be far smaller than the injector size \citep{RubioRubio2013,EggersVillermaux2008}. 

For Newtonian liquids injected vertically downwards, the dynamics exhibit a sharp transition between a steady jetting regime and a dripping regime in which drops detach directly from the nozzle. 
A long filament develops only when the imposed flow rate exceeds a critical threshold $Q_c(\rho,\nu,\sigma,g,R)$ (or, in nondimensional form, a critical Weber number $\We_c(\Bo,\Gamma)$), which depends on density $\rho$, kinematic viscosity $\nu$, surface tension $\sigma$, gravity $g$, and nozzle radius $R$ \citep{RubioRubio2013}. 
Experiments also show a hysteretic behaviour, for which the critical flow rate for the disappearance of an initially long jet (jetting to dripping) is lower than that required to form a long jet when increasing the flow rate starting from dripping \citep{clanet1999transition,ambravaneswaran2004dripping}. 
Since the minimum achievable drop size for a given liquid and injector decreases with the critical flow rate, the jetting-to-dripping boundary is the relevant stability limit when the objective is to sustain the thinnest possible steady threads while minimizing injector complexity \citep{RubioRubio2013}. 

From a theoretical viewpoint, early attempts to rationalize the jetting--dripping transition invoked local (parallel-flow) spatio-temporal stability analyses of capillary jets. 
For a uniform cylindrical jet, \citet{leib1986generation} showed that there exists a critical flow rate below which the flow becomes absolutely unstable and a spatial analysis is not well posed \citep[see also][]{keller1973spatial}; in this framework the transition is often discussed in terms of convective versus absolute amplification. 
A useful point of contact with the present study is the local spatio-temporal analysis of \citet{alhushaybari2019convective}, who examine convective and absolute instability of a viscoelastic jet falling under gravity by deriving a dispersion relation for axisymmetric travelling-wave modes about a slowly varying slender base state and then applying standard pinch-point criteria in the complex wavenumber and frequency planes. 
The main strength of such local approaches is their diagnostic clarity: they provide a direct regime classification in terms of convective versus absolute instability and enable rapid parameter sweeps without solving a global eigenvalue problem. 
Their limitation in the strongly stretched regime is that the base flow is treated as quasi-parallel over the perturbation scale, whereas in gravitationally stretched jets the radius and velocity can vary on the same axial scale as the instability core and the nozzle boundary conditions play a decisive role in selecting the observed global oscillator. 
In that case, a global formulation retains the non-parallel base state and boundary conditions and yields discrete global eigenpairs with onset frequencies, together with structural-sensitivity diagnostics that quantify where and through which couplings the instability is selected. 
This perspective explains why local wave-train predictions can become, at best, order-of-magnitude estimates of the experimentally observed thresholds in gravitationally stretched jets \citep{RubioRubio2013}, and it motivates the use of global stability analyses when quantitative marginal curves and receptivity mechanisms are sought.

Several studies have therefore incorporated non-parallel and global effects in the dripping--jetting problem. 
\citet{schulkes1994evolution} obtained a dripping--jetting transition in the ideal-flow limit; for inertia-dominated jets, \citet{le1997global} identified a global mechanism preventing the formation of a slender jet at low exit velocities; and \citet{senchenko2004shape} used multiple-scale methods to quantify a stabilizing influence of gravity on capillary perturbations. 
In addition, viscous relaxation of the exit velocity profile can play an important role in low-viscosity liquids and long injectors, improving agreement between theory and experiments when incorporated appropriately \citep{sevilla2011effect}. 
In the gravitationally stretched regime of interest here, a key step was taken by \citet{sauter2005stability}, who combined experiments and a global linear analysis for very viscous jets and identified the self-excited global mode associated with the onset of unsteadiness. 
Later, \citet{RubioRubio2013} demonstrated that one-dimensional slender-jet equations can predict not only the critical flow rate but also the oscillation frequency at onset over wide ranges of viscosity and nozzle diameter, provided that the exact interfacial curvature is retained \citep[see also][]{EggersDupont1994,garcia1994one}. 
A central conclusion of that work is that axial curvature, which is already recognized as stabilizing in the classical Plateau--Rayleigh setting, plays an essential stabilizing role in strongly stretched jets, especially for large Bond numbers, thereby reshaping the stability boundary in practically relevant non-slender configurations \citep{RubioRubio2013}. 

Parallel to these developments, polymer solutions and other weakly elastic liquids are known to exhibit markedly different thinning and breakup dynamics relative to Newtonian liquids. 
Even at dilute concentrations, extensional stresses can delay pinch-off, promote elastocapillary thinning, and generate beads-on-a-string morphologies; these regimes have been captured quantitatively with one-dimensional free-surface models coupled to Oldroyd--B or Giesekus constitutive laws, enabling inference of extensional rheology from the evolution of jets and filaments \citep{Ardekani2010}. 
Moreover, dispensing and jetting experiments in polymer solutions reveal additional dynamical regimes and transitions, including oscillatory behaviours, that depend sensitively on elasticity and on the interplay between gravity, capillarity and extensional stress \citep{clasen2009gobbling}. 
Complementing these experimental observations, local spatio-temporal analyses of viscoelastic jets under gravity have also quantified how elasticity can shift convective and absolute instability boundaries and thereby modify the conditions for unsteadiness~\citep{alhushaybari2019convective}, as pointed out above. 
Together, these results suggest that the critical flow rate and the onset mechanism in gravitationally stretched jets should be substantially modified by viscoelasticity in experimentally accessible parameter ranges.

A key open direction, therefore, is to merge these two threads: develop a predictive global-stability framework for jets that are simultaneously (i) strongly stretched by gravity and (ii) governed by viscoelastic constitutive dynamics. 
Beyond determining how the marginal conditions $\We_c(\Bo)$ and the associated onset frequency shift with $(\beta,\De)$, a structural-sensitivity analysis can localize the instability mechanism, identify which equation--variable couplings are most responsible for eigenvalue drift, and provide a principled bridge between global stability and receptivity/forcing considerations in open flows \citep{GiannettiLuchini2007,marquet2015identifying}. 
In the present setting this is particularly appealing, since gravitational stretching produces a spatially developing base state whose instability core may be localized near the inlet while the observable dynamics extend far downstream; direct/adjoint diagnostics offer a compact route to isolate the effective feedback region and to interpret how polymeric tension modifies it.

More recent work has broadened the Newtonian jetting picture by explicitly targeting mechanisms that sit beyond the onset eigenvalue itself, including geometric confinement, noise-driven amplification, and forced receptivity. In particular, \citet{martinez2018nonlinear} extended the global framework of \citet{sauter2005stability,RubioRubio2013} to axially confined jets whose lower end interacts with a bath, and showed that the confinement length $L$ is an additional control parameter that shifts the critical flow rate and can even introduce an intermediate nonlinear regime of sustained limit-cycle oscillations without breakup, denoted oscillatory jetting. In the globally stable jetting regime, \citet{le2017capillary} instead focused on how perturbations grow convectively and quantified optimal spatial gains using a WKBJ approach for both nozzle forcing and distributed background noise, thereby linking breakup length, dominant wavelength and droplet size to a noise-amplification mechanism rather than to a discrete unstable global mode. These questions naturally connect to non-modal viewpoints, and \citet{shukla2020frequency} pursued this direction by combining nonlinear simulations with a global resolvent analysis of gravitationally stretched jets in the stable regime, extracting the optimal forcing frequency and spatial response while highlighting a pronounced dependence on forcing amplitude when gravity-induced non-parallelism is active. Finally, \citet{sun2024global} incorporated insoluble surfactants and Marangoni stresses into a global model and carried out both global stability and global resolvent analyses, showing that surfactants and gravity stabilize the base state while gravitational stretching can shift the optimal forcing frequency away from the intrinsic one for stable jets, whereas near criticality a resonance-like response locks the optimal forcing to the leading global frequency. Collectively, these studies emphasize that, in stretched jets, the observed unsteadiness can reflect a competition between intrinsic global modes and strong convective or forced amplification shaped by non-parallelism, boundary conditions, and additional interfacial physics, which motivates treating frequency selection, receptivity, and sensitivity within a global framework.

The paper is structured as follows:
\S~\ref{sec:mathematical_description} formulates the one-dimensional slender-jet model for gravitationally stretched viscoelastic jets, \S~\ref{sec:linearised_eqs} derives the linearized equations about the steady base state and summarizes the global eigenvalue formulation, the discrete adjoint problem, and the wavemaker/endogeneity diagnostics used to interpret the instability mechanism. 
\S~\ref{sec:case_studies} presents selected case studies, \S~\ref{sec:wavemaker} uses wavemaker and endogeneity decompositions to identify the regions and couplings responsible for eigenvalue selection, and \S~\ref{sec:marginal} quantifies how viscoelasticity shifts marginal conditions and onset frequencies. We conclude in~\S~\ref{sec:conclusions} with a brief discussion of implications and open directions.


\section{Mathematical description}
\label{sec:mathematical_description}
We consider an incompressible axisymmetric slender viscoelastic jet with radius $r(z,t)$ and mean axial velocity $u(z,t)$. The solvent has constant density $\rho$ and viscosity $\mu_s$, whereas the polymer is characterised by its viscosity $\mu_p$ and relaxation time $\lambda$. The liquid-air surface tension coefficient is $\sigma$, assumed constant as well. The description of the jet dynamics combines the leading-order mass and momentum equations, derived originally by~\citet{EggersDupont1994}, augmented by the extra stresses due to the presence of the polymer~\citep{Ardekani2010}, given by
\begin{equation}
\partial_t(r^2) + \partial_z(u r^2)=0,
\label{eq:mass}
\end{equation}
and 
\begin{equation}
\rho \left( \partial_t u + u\,\partial_z u \right)
= 
\rho g - \sigma \partial_z \mathcal{C}
+ \frac{3 \mu_s }{r^2}\partial_z\!\left(r^2\,\partial_z u\right)
+ 
\frac{1}{r^2}\partial_z\!\left(r^2\,T\right),
\label{eq:mom}
\end{equation}
where $g$ is the gravitational acceleration, $\mathcal{C}$ is the full curvature, 
\begin{equation}
\mathcal{C} = \frac{1}{r}\left(1+r_z^2\right)^{-1/2}
\!\!\! - 
r_{zz}\left(1+r_z^2\right)^{-3/2},
\label{eq:curvature}
\end{equation}
and $T = \tau_{zz} - \tau_{rr}$ is the polymeric tensile stress, written in terms of the diagonal components of the polymeric extra-stress tensor $\bm{\tau}$. The evolution of $\bm{\tau}$ is assumed to be dictated by the Giesekus constitutive model 
\begin{equation}
\label{eq:dimGiesekus}
\bm{\tau} + \lambda\,\overset{\triangledown}{\bm{\tau}} + \frac{\alpha\lambda}{\mu_p}\,\bm{\tau}\cdot\bm{\tau}
= 2\mu_p\,\bm{D},
\end{equation}
with $\bm{D}=\tfrac12(\nabla\bm{v}+\nabla\bm{v}^T)$, $\overset{\triangledown}{\bm{\tau}}$ the upper-convected derivative, and $\alpha$ a dimensionless constant which. Note that setting $\alpha=0$ recovers the Oldroyd-B model, which will be the choice for the rest of this paper. The diagonal components of~\eqref{eq:dimGiesekus} read
\begin{align}
\lambda\Big(\partial_t\tau_{zz} + u\,\partial_z\tau_{zz} - 2\,u_z\,\tau_{zz}\Big)
+ \tau_{zz} + \frac{\alpha\lambda}{\mu_p}\,\tau_{zz}^2
&= 2\mu_p\,u_z,
\label{eq:tauzz}\\
\lambda\Big(\partial_t\tau_{rr} + u\,\partial_z\tau_{rr} + u_z\,\tau_{rr}\Big)
+ \tau_{rr} + \frac{\alpha \lambda}{\mu_p}\,\tau_{rr}^2
&= -\mu_p \,u_z.
\label{eq:taurr}
\end{align}

The mathematical description is completed upon boundary conditions, namely,
\begin{equation}
\label{eq:bc_inlet}
r(0,t)-R \;=\; u(0,t)-\frac{Q}{\pi R^2} \;=\; \tau_{zz}(0,t) \;=\; \tau_{rr}(0,t) \;=\; 0,
\end{equation}
where $R$ is the nozzle radius and $Q$ is the prescribed flow rate. The assumption of fully relaxed polymer at the exit, $\tau_{zz}(0)=\tau_{rr}(0)=0$, provides a minimal and internally consistent closure for a free-jet model posed only for $z\ge 0$. It is appropriate when the residence time in the injector is long compared with the relaxation time $\lambda$ and when upstream deformation rates are weak. At the same time, it is not expected to be universally accurate in experiments, since polymer solutions may exit the injector with a non-negligible elastic memory of the internal shear and extension history, as evidenced by the classical die-swell (Barus) effect. In particular, extrudate swell in dilute polymer solutions is driven primarily by shear-generated normal stresses, notably the first normal-stress difference, rather than by a stress-free state at the die exit \citep[e.g.][]{mitsoulis2010extrudate}, and elasticity is known to influence jetting--dripping thresholds in related jetting configurations \citep{montanero2008viscoelastic}. 

In the present work we nevertheless adopt \eqref{eq:bc_inlet} as our inlet stress condition. The reason is that any non-zero specification of $\tau_{zz}(0)$ and $\tau_{rr}(0)$ within a reduced free-surface jet model is necessarily an effective closure unless the internal nozzle flow and die-exit region are modelled explicitly. Asymptotic inlet stress relations derived from the outer jet equations, as in local slender analyses of viscoelastic jets under gravity \citep{alhushaybari2019convective}, are mathematically consistent with the idealized upstream state implicit in the expansion, but they do not encode the dependence on injector geometry, entrance effects, or residence-time history that controls die swell in practice. Absent such a nozzle-level description, prescribing finite inlet stresses would introduce additional parameters whose values are not constrained by the present model and would therefore remain essentially arbitrary. The relaxed inlet condition \eqref{eq:bc_inlet} should thus be interpreted as a baseline choice that isolates the role of viscoelasticity in the downstream stretched-jet dynamics. Incorporating a coupled nozzle--jet model or a calibrated inlet-stress closure based on independent measurements of die swell and upstream rheology is a natural extension of the present framework and is deferred to future work.

Lastly, initial conditions are also needed for the temporal marching of the model. We shall describe them below if needed.

\subsection{Choice of scales and non-dimensionalisation}

Following \citet{RubioRubio2013}, we take the gravito--capillary scales 
\begin{equation}
\label{eq:scales}
\ell_\sigma = \left(\frac{\sigma}{\rho g}\right)^{1/2}, \quad
U_\sigma=\sqrt{g\ell_\sigma}, 
\quad
T_\sigma=\sqrt{\ell_\sigma/g},
\quad 
\tau_c = \mu_p U_\sigma/\ell_\sigma,
\end{equation}
as relevant scales for length, velocity, time, and polymer stress, respectively. With these choices, the dimensionless model reads
\begin{equation}
\partial_t(r^2) + \partial_z(u r^2)=0.
\label{eq:mass_nondim}
\end{equation}

\begin{equation}
\partial_t u + u\,\partial_z u
= 1 - \partial_z \mathcal{C}
+ \frac{\Gamma}{r^2}\partial_z\!\left(r^2\,\partial_z u\right)
+ \frac{\Gamma (1-\beta)}{3\beta r^2}\partial_z\!\left(r^2\,T\right),
\label{eq:mom_nondim}
\end{equation}

\begin{align}
\De\Big(\partial_t\tau_{zz} + u\,\partial_z\tau_{zz} - 2\,u_z\,\tau_{zz}\Big)
+ \tau_{zz} + {\alpha\De}\,\tau_{zz}^2
&= 2 u_z,
\label{eq:tauzz_nondim}\\
\De\Big(\partial_t\tau_{rr} + u\,\partial_z\tau_{rr} + u_z\,\tau_{rr}\Big)
+ \tau_{rr} + {\alpha\De} \,\tau_{rr}^2
&= -\,u_z,
\label{eq:taurr_nondim}
\end{align}
where each dependent variable should be understood as scaled with~\eqref{eq:scales}, i.e., $z \to z/\ell_\sigma$, $u \to u/U_\sigma$, $t \to t/T_\sigma$, $\tau_{ij} \to \tau_{ij}/\tau_c$. In these units, the dimensionless counterpart of boundary conditions~\eqref{eq:bc_inlet} are prescribed by
\begin{equation}
r(0,t)=\Bo^{1/2}, \quad u(0,t) = \We^{1/2}\Bo^{-1/4}, \quad 
\tau_{zz}(0,t) = \tau_{rr}(0,t) = 0.
\label{eq:inletBC_dimless}
\end{equation}
Suitable {\em outlet} conditions must also be prescribed downstream in order to numerically integrate these equations. A possible choice is a jet with zero axial curvature, i.e., $r_z = r_{zz} = 0$, at $z = L \gg 1$, as done for instance by~\cite{shukla2020frequency}.
Above, $\Bo = \rho g R^2/\sigma$ is the Bond number, $\We = \rho U^2 R/\sigma$ is the Weber number based on the mean exit speed $U = Q/(\pi R^2)$, $\Gamma = {3\nu_s}/{\sqrt{g\ell_\sigma^3}}$ is the Kapitza number based on the solvent kinematic viscosity $\nu_s = \mu_s/\rho$, $\De=\lambda/T_\sigma = \lambda\sqrt{g/\ell_\sigma}$ is the Deborah number, and $\beta = \mu_s/(\mu_s + \mu_p)$ is the solvent-to-total viscosity ratio. Finally, note that the Newtonian limit is recovered for $1-\beta = \De = 0$.

\subsection{A note on rheological parameters}

In order to make sure that the rheological parameters, mostly $\De$ and $\beta$, remain within a realisable experimental range, we compute their values for a couple of examples from~\cite{clasen2012dispensing}. In their ``map of misery'' study they prepare (i) a polystyrene solution (PS) in diethyl phthalate (DEP) and (ii) a polyisobutylene solution (PIB) in pristane (an isoparaffinic hydrocarbon), and report the corresponding material properties $(\rho,\sigma,\lambda)$. Using scales~\eqref{eq:scales}, these two canonical polymer solutions map to $\De\simeq 1.68$ (PS in DEP: $\rho=1118~\mathrm{kg/m^3}$, $\sigma=37.5~\mathrm{mN/m}$, $\lambda=0.023~\mathrm{s}$) and $\De\simeq 4.22$ (PIB in pristane: $\rho=793.7~\mathrm{kg/m^3}$, $\sigma=24.9~\mathrm{mN/m}$, $\lambda=0.057~\mathrm{s}$), directly justifying a focus on moderate elasticity $\De=O(1)$. Finally, their third viscoelastic reference liquid is a Boger fluid, namely a (nearly) constant-shear-viscosity elastic liquid obtained by dissolving a trace amount of high-molecular-weight PIB in a high-viscosity, low-molecular-weight PIB matrix; this illustrates in practice how elasticity can be tuned largely independently of the shear viscosity level. In our notation, this corresponds to the experimentally accessible regime where the ``solvent-like'' contribution dominates the shear viscosity (i.e.\ $\beta$ closer to $1$), while the polymeric relaxation time (and thus $\De$) can be made very large if desired.

\section{The linearised equations}
\label{sec:linearised_eqs}
We assume the usual normal-mode decomposition for each fluid variable $\bm{q} = [r, u, \tau_{zz}, \tau_{rr}]^\mathrm{T}$ about a steady base,
\begin{equation}
    \label{eq:ansatz_normal_mode}
    \bm{q}(z,t) = \bm{q}_0(z) + \epsilon \hat{\bm{q}}(z)\mathrm{e}^{\omega t},
\end{equation}
where $\epsilon \ll 1$ is an arbitrarily small number, $\bm{q}_0(z)$ is the steady solution to equations~\eqref{eq:mom_nondim}--\eqref{eq:taurr_nondim}, $\omega=\omega_r +\mathrm{i}\omega_i$ is a complex growth rate whose real and imaginary parts represent the temporal amplification rate and oscillation frequency, respectively, and $\hat{\bm{q}}$ are (direct) eigenfunctions. Substituting the ansatz~\eqref{eq:ansatz_normal_mode} into~\eqref{eq:mom_nondim}--\eqref{eq:taurr_nondim} and collecting $O(\epsilon)$ terms yields a linear system that can be written in compact form as the generalised eigenvalue problem
    \begin{equation}
        \label{eq:direct_gevp}
        \left( \mathcal{A} - \omega \mathcal{B} \right) \hat{\bm{q}} = \bm{0},
    \end{equation}
where $\mathcal{A}$ and $\mathcal{B}$ are linear differential operators encompassing boundary conditions. We employ a standard Chebyshev collocation technique in order to discretize~\eqref{eq:direct_gevp} which in turn yields the matrix pencil $(\mathsfbi{A} - \omega \mathsfbi{B}) \bm{\tilde{q}} = 0$, where $\mathsfbi{A}, \mathsfbi{B}$, and $\tilde{\bm{q}}$ represent the discretised counterparts, whose spectrum $\omega$ determines the global temporal stability of the spatially varying base flow. This formulation, accounting for full curvature~\eqref{eq:curvature}, was employed for the first time in the framework for Newtonian stretched jets~\citep{RubioRubio2013}. Because the base state varies significantly with $z$ under gravitational acceleration and capillary thinning, local parallel stability theory is insufficient to capture the onset of self-excited jet oscillations. Thus, a global approach is the correct strategy to predict stability thresholds and oscillation frequencies in the jetting regime. In the discretized setting, we define a weighted inner product consistent with the spectral collocation quadrature,
\begin{equation}
    \label{eq:inner_product_Q}
    \langle \bm{a},\bm{b}\rangle_Q \equiv \bm{a}^\mathrm{H}\,\mathsfbi{Q}\,\bm{b},
\end{equation}
where $(\cdot)^\mathrm{H}$ denotes the conjugate transpose and $\mathsfbi{Q}$ is a Hermitian positive-definite weight matrix assembled from the collocation weights (see, for instance, \cite{Trefethen2000}). Direct eigenfunctions are normalised such that 
    \begin{equation}
        \langle \tilde{\bm{q}} , \tilde{\bm{q}} \rangle_Q = 1.
    \end{equation}  
To quantify the sensitivity of the global modes to localized modifications of the base flow and/or of the linear operator, we also introduce the corresponding adjoint (left) eigenfunctions. From a mathematical standpoint, the adjoint eigenfunction provides the receptivity kernel of the global oscillator, i.e., it weights external forcing and initial perturbations in the energy inner product, thereby filtering generic disturbances into the component that most efficiently excites the selected direct mode and pinpointing the region(s) of strongest feedback sensitivity. The (discrete) adjoint eigenfunctions $\tilde{\bm{q}}^{\dagger}$ are then defined as solutions of the $Q$-adjoint generalized eigenvalue problem
\begin{equation}
    \label{eq:adjoint_gevp}
    \left(\mathsfbi{A}^{\dagger}-\bar{\omega}\,\mathsfbi{B}^{\dagger}\right)\tilde{\bm{q}}^{\dagger}=\bm{0}, 
    \quad     \mathsfbi{A}^{\dagger}=\mathsfbi{Q}^{-1}\mathsfbi{A}^{\mathrm{H}}\mathsfbi{Q},\quad    \mathsfbi{B}^{\dagger}=\mathsfbi{Q}^{-1}\mathsfbi{B}^{\mathrm{H}}\mathsfbi{Q},
\end{equation}
and we impose the standard bi-orthogonality (normalization) condition
\begin{equation}
    \label{eq:biorthogonality}
    \left\langle \tilde{\bm{q}}^{\dagger},\mathsfbi{B}\,\tilde{\bm{q}} \right\rangle_Q
    = \tilde{\bm{q}}^{\dagger \mathrm{H}}\mathsfbi{Q}\,\mathsfbi{B}\,\tilde{\bm{q}}
    =1,
\end{equation}
which enables a direct projection of perturbations onto individual global modes. 

Building on this framework, we characterize the spatial structure of the global instability using two complementary diagnostics based on the sensitivity of eigenvalues with respect to localized changes of the operator $\mathcal{A}$. First, the \emph{wavemaker} (or structural sensitivity) field measures the sensitivity of $\omega$ to a localized \emph{feedback} perturbation of the operator and is typically constructed from the pointwise overlap of the direct and adjoint eigenfunctions; it highlights the region where the instability mechanism is most receptive to modifications of the local coupling between state variables, and is therefore a proxy for the core of the global feedback loop~\citep{GiannettiLuchini2007}. It is defined as the real scalar
    \begin{equation}
        S(z) = \norm{  \mathsfbi{Q} \tilde{\bm{q}}^\dagger } \,  \norm{\tilde{\bm{q}}},   
    \end{equation}
which corresponds to the Frobenius norm of the structural-sensitivity tensor
    \begin{equation}
        \mathsfbi{S} = \mathsfbi{Q} \tilde{\bm{q}}^\dagger \otimes  \tilde{\bm{q}}.
    \end{equation}
Additionally, the momentum-restricted wavemaker is defined as 
\begin{equation}
        S_\mathrm{mom}(z) = \norm{  \mathsfbi{U} ( \mathsfbi{Q} \tilde{\bm{q}}^\dagger ) } \,  \norm{ \mathsfbi{U} \tilde{\bm{q}}},   
    \end{equation}
where $\mathsfbi{U}$ is a projection matrix that extracts the subset of state components associated to the momentum equation. 
Second, the \emph{endogeneity} $E = E_r + \mathrm{i} E_i$, defined as 
\begin{equation}
    E(z) = \tilde{\bm{q}}^\dagger  \cdot \left(\mathsfbi{A} \tilde{\bm{q}} \right), 
\end{equation}
where the dot indicates scalar multiplication of two vectors, provides a local budget for eigenvalue selection by quantifying the pointwise contribution of the unperturbed operator acting on the direct mode, weighted by the adjoint mode. In its discrete form, it yields a spatial density whose integral recovers the eigenvalue, namely, 
\begin{equation}
    \int_0^L E(z) \, \mathrm{d}z = \omega,    
\end{equation}
and whose real and imaginary parts can be interpreted as contributions to growth-rate and frequency selection, respectively~\citep{marquet2015identifying}. Together, wavemaker and endogeneity allow us to distinguish where the flow is \emph{most sensitive} to localized perturbations (wavemaker) from where the eigenvalue is \emph{intrinsically produced} by the dynamics of the unperturbed base state (endogeneity), thereby providing a sharp picture of the regions controlling global instability in stretched Newtonian and viscoelastic jets.


\section{Selected case studies}
\label{sec:case_studies}
This section presents a set of representative case studies designed to (i) benchmark the present base-flow and global-stability formulation against established Newtonian results and (ii) illustrate, in a controlled manner, how weak elasticity modifies both the onset and the structure of the global oscillatory instability. We focus on parameter combinations that lie close to marginality so that (a) the leading eigenpair is cleanly separated from the rest of the spectrum and (b) small changes in rheology produce measurable eigenvalue drift. The first case reproduces the nearly marginal Newtonian jetting configuration of~\citet{RubioRubio2013} and serves as a reference for our numerical implementation (base state, spectrum, and eigenfunctions), after which we introduce viscoelasticity at fixed viscosity ratio to quantify the role of polymeric tension in the linear dynamics. The second case is a nearly marginal viscoelastic jet for which we validate the global-mode prediction by direct time marching of the non-linear one-dimensional equations under a weak, transient inlet perturbation. Together, these two cases establish the reliability of the global framework and provide the building blocks for the mechanistic diagnostics developed below (wavemaker, endogeneity, and their decomposition), as well as for the parametric trends in marginal conditions discussed subsequently. 

\begin{figure}
    \centering    \includegraphics[width=0.99\linewidth]{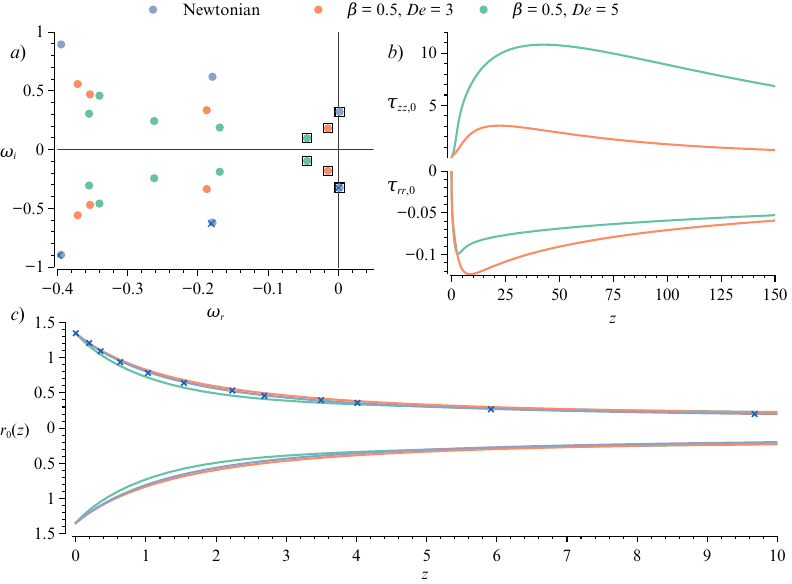}
    \caption{a) Direct spectrum, where hollowed out squares mark the leading eigenvalues, b) spatial evolution of the base polymeric stresses, and c) base jet radius, for $\We = 0.003$, $\Bo = 1.81$, $\Gamma = 5.83$ and selected rheology parameters (see legend). Dark blue crosses in a) and c) are Newtonian results from~\cite{RubioRubio2013}, included for comparison.}
    \label{fig:fig1}
\end{figure}

\subsection{The case $\We = 3\times 10^{-3}$, $\Bo = 1.81$, $\Gamma = 5.83$}

This case serves as a benchmark for our base-flow and linear-stability results. In the Newtonian limit, $1-\beta=\De=0$,~\citet{RubioRubio2013} identified this parameter combination as nearly marginal, i.e., the leading global mode has vanishing growth rate, $\omega_r\simeq 0$, while oscillating at $\omega_i\simeq 0.325$, thereby fixing the critical flow rate (Weber number) for a prescribed liquid (Kapitza number) and nozzle diameter (Bond number) at the onset of self-sustained jet oscillations in the jetting--dripping transition. Figure~\ref{fig:fig1}(a,c) shows that our global spectrum and base-state radius $r_0(z)$ accurately reproduce the reference Newtonian results, including the marginal leading complex-conjugate pair and the weakly non-parallel thinning of the jet. We then extend this benchmark by introducing viscoelasticity at fixed solvent-to-total viscosity ratio $\beta=0.5$, considering $\De=3$ and $\De=5$. The spectrum in figure~\ref{fig:fig1}(a) shows that order-unity (and larger) Deborah numbers are required for the leading eigenvalue to drift appreciably, meaning that increasing $\De$ shifts the dominant pair towards more negative $\omega_r$ and simultaneously reduces $|\omega_i|$, indicating a stabilizing influence and a lower global oscillation frequency at fixed $(\We,\Bo,\Gamma)$. Consistently, the base state develops a sizeable polymeric tensile resistance $T_0=\tau_{zz,0}-\tau_{rr,0}\simeq \tau_{zz,0}$, with $\tau_{rr,0}$ remaining comparatively small and $\tau_{zz,0}$ reaching $O(1)$ values for $\De=3$ and $O(10)$ values for $\De=5$ over an extended downstream region (figure~\ref{fig:fig1}b). Physically, this accumulated extensional stress provides an additional streamwise tension that competes with capillary-driven thinning, thereby lowering the global growth rate and delaying the onset of a self-excited oscillatory state (at fixed $\We$). At the same time, the base radius profile remains close to the Newtonian one (figure~\ref{fig:fig1}c), with only modest quantitative changes near the nozzle, which is consistent with viscoelasticity acting primarily through an added tensile stress pathway rather than through a gross reconfiguration of the mean jet geometry in this parameter regime.

\subsection{The case $\We = 7\times 10^{-4}$, $\Bo = 1.8$, $\Gamma = 5.8$}

\begin{figure}
    \centering
    \includegraphics[width=0.99\linewidth]{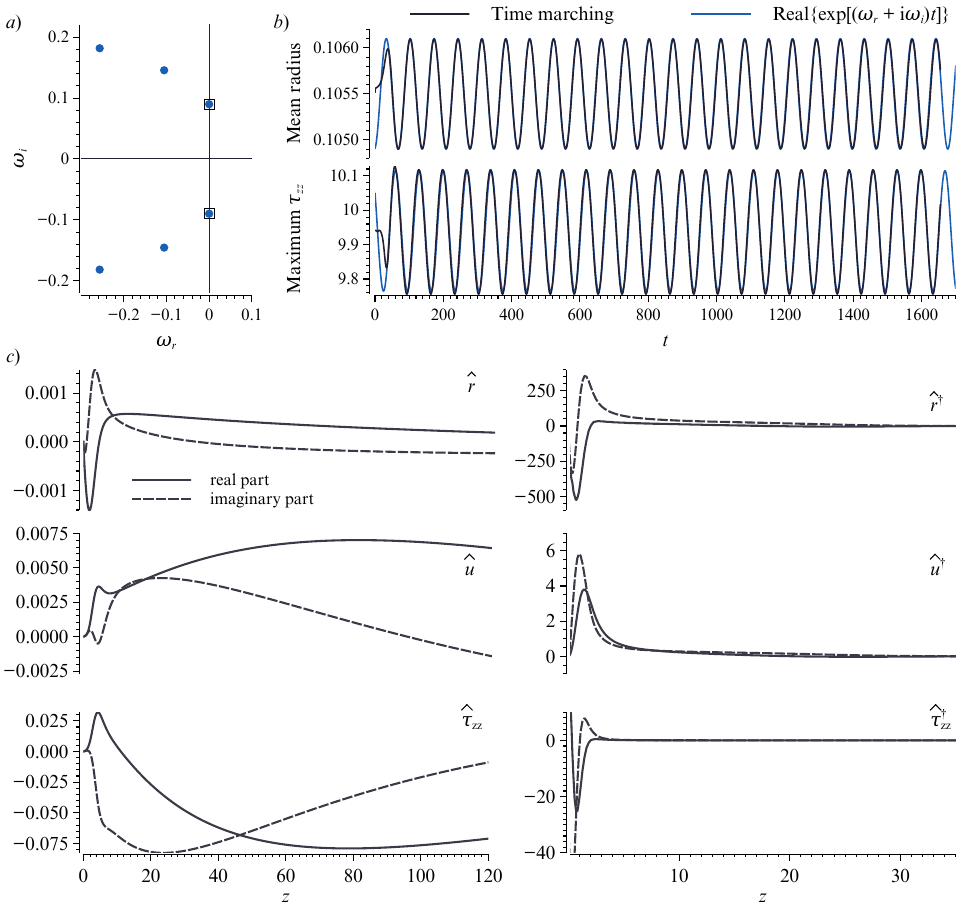}
    \caption{Marginally stable jet for $\We = 7\times 10^{-4}$, $\Bo = 1.8$, $\Gamma = 5.8$, $\beta = 0.5$, $\De = 5$. a) Direct spectrum, for which the leading mode (marked with a square) is $\omega \simeq -1.211\times 10^{-6} \pm 0.08984\mathrm{i}$. b) Comparison between time marching and the leading linear eigenmode dynamics. c) and d) Direct and adjoint normalised eigenfunctions.}
    \label{fig:fig2}
\end{figure}

This case is chosen to be nearly marginal (weakly damped) in the viscoelastic regime, with a leading complex-conjugate pair
$
\omega \simeq -1.211\times 10^{-6}\pm 0.08984\,\mathrm{i},
$
so that the growth rate is essentially zero on the time scales of interest while a well-defined global oscillation frequency is selected. Figure~\ref{fig:fig2}a) shows the direct spectrum, where the dominant pair lies closest to the imaginary axis and is well separated from the remaining (more strongly damped) eigenvalues, indicating that the long-time linear dynamics are governed by a single global Hopf-type mode rather than by a broadband convective response, as in the Newtonian case. To validate this prediction in the fully non-linear time-dependent equations, we perform a direct time-marching of~\eqref{eq:mass_nondim}--\eqref{eq:taurr_nondim} using the steady base state $\bm{q}_0(z)$ as initial condition, and we trigger the response via a small, transient inlet perturbation of the velocity,
\begin{equation}
u(0,t) = \We^{1/2}\Bo^{-1/4}\left[1+\delta\,\mathrm{e}^{-a(t-t^*)^2}\,\mathcal{H}(t-t^*)\right],
\end{equation}
where $\mathcal{H}$ is the Heaviside function and $\delta=0.01$ is sufficiently small for the ensuing evolution to remain in the linear regime and qualitatively independent of the forcing amplitude. The parameters $a=10$ and $t^*=10^{-3}$ control the width and onset of the notch and are selected to facilitate numerical start-up while providing a localised impulse-like excitation of the global dynamics. As shown in figure~\ref{fig:fig2}b), both the mean radius and the maximum axial polymer stress $\tau_{zz}$ extracted from the time marching rapidly lock onto the linear prediction. After a short transient, the signals oscillate at the eigenfrequency $\omega_i$ with an envelope consistent with $\exp(\omega_r t)$ (here essentially constant owing to $|\omega_r|\ll 1$), thereby confirming the global-mode interpretation. Finally, the spatial structure of the associated eigenfunctions (figure~\ref{fig:fig2}c) further supports this picture. The direct mode exhibits an extended downstream support (consistent with a convective redistribution of perturbations along the stretched jet), whereas the adjoint mode is strongly localized near the inlet, indicating that the receptivity and mode selection are controlled predominantly by the near-nozzle region. This marked upstream localization of the adjoint fields underscores that small perturbations imposed at, or near, the inlet are the most effective at exciting the observed oscillatory response.

\section{Wavemaker and endogeneity: the origin of instability}
\label{sec:wavemaker}

\begin{figure}
    \centering
    \includegraphics[width=0.95\linewidth]{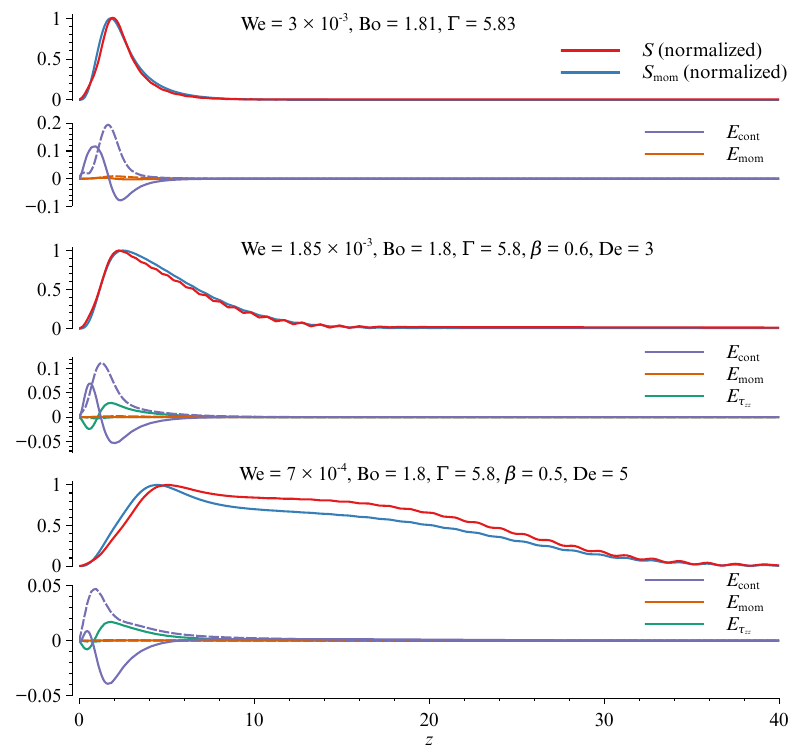}
    \caption{Wavemaker and endogeneity spatial distributions for three separate cases: $(\We, \Bo, \Gamma) = (3\times 10^{-3},1.81, 5.83)$ (Newtonian), $(\We, \Bo, \Gamma, \beta, \De) = (1.85 \times 10^{-3}, 1.8, 5.8, 0.6, 3)$, and $(\We, \Bo, \Gamma, \beta, \De) = (7 \times 10^{-4}, 1.8, 5.8, 0.5, 5)$. For endogeneity computations, solid lines indicate real part, and dashed lines indicate imaginary part.}
    \label{fig:fig3}
\end{figure}

Figure~\ref{fig:fig3} compares wavemaker and endogeneity diagnostics for three nearly marginal configurations, namely, a Newtonian reference case (top panels) and two viscoelastic cases of increasing elasticity (middle and bottom panels). In all three cases the parameters are chosen close to the onset of global instability, so that the leading eigenpair is weakly damped (or nearly neutral) and the associated spatial diagnostics can be interpreted as the footprint of the global feedback loop. Throughout, solid and dashed lines denote the real and imaginary parts of the endogeneity, respectively; accordingly, $\Re(E)$ quantifies the local contribution to the global growth/decay rate, while $\Im(E)$ quantifies the local contribution to the global oscillation frequency. For the Newtonian case, the normalized wavemaker $S(z)$ is sharply localised near the inlet and decays rapidly over a few nozzle radii, while $S_{\mathrm{mom}}(z)$ is nearly indistinguishable from $S(z)$, indicating that the feedback region is compact and that restricting the structural-sensitivity tensor to the $u$-row captures essentially the same spatial support. Consistently, the endogeneity is dominated by the continuity/capillary pathway, i.e., $E_{\mathrm{cont}}$ is $O(10^{-1})$ locally and displays the expected sign structure in its real part (a positive lobe followed by a compensating negative lobe), so that the net integrated contribution remains small in the marginal limit, whereas $E_{\mathrm{mom}}$ is negligible in both real and imaginary parts. The imaginary component $\Im(E_{\mathrm{cont}})$ is also confined to the upstream region, supporting the interpretation that the onset frequency is set primarily by the near-nozzle feedback core.

The viscoelastic cases retain the same qualitative global-oscillator signature but reveal a progressive and physically meaningful change in the spatial organization of the sensitivity as elasticity is increased. For the intermediate case ($\beta=0.6$, $\De=3$, middle panels), the wavemaker remains upstream-peaked but its decay is already noticeably slower than in the Newtonian limit, producing a downstream tail that extends over a significantly larger fraction of the domain. This trend is further amplified in the more elastic case ($\beta=0.5$, $\De=5$, bottom panels), for which $S(z)$ becomes markedly more distributed and remains non-negligible over an extended downstream interval. In other words, one of the most robust effects of introducing viscoelastic dynamics in a gravitationally stretched jet is not merely a shift of the leading eigenvalue, but a pronounced preading of the feedback-sensitive region. The instability is still selected upstream, yet the portion of the jet that participates in the global feedback loop grows with $\De$. This behaviour is consistent with the fact that polymeric stresses are advected and relax over a finite time scale, so that elastic memory couples distant axial locations through the transport of tensile stress. At the same time, the separation between $S(z)$ and $S_{\mathrm{mom}}(z)$ becomes increasingly apparent as elasticity grows (most clearly in the bottom case), for which $S(z)$ exceeds $S_{\mathrm{mom}}(z)$ over much of the downstream tail, indicating that a momentum-row restriction underestimates the full structural sensitivity once viscoelastic degrees of freedom are active and confirming that the  {\em extra} sensitivity is carried by the stress subspace rather than by the Newtonian momentum balance alone.

The endogeneity decomposition confirms this interpretation while highlighting that the dominant balance remains localized closer to the inlet than the wavemaker itself. In both viscoelastic cases, $E_{\mathrm{mom}}$ remains small, but an additional contribution associated with polymeric tension emerges through $E_{\tau_{zz}}$. This elastic contribution overlaps with the region where $E_{\mathrm{cont}}$ is active and provides a clear mechanism for eigenvalue selection. The onset results from a coupled capillary--elastic feedback pathway in which kinematic/capillary thinning is opposed by an additional tensile-stress response. The real parts of $E_{\mathrm{cont}}$ and $E_{\tau_{zz}}$ exhibit comparable magnitudes in the upstream region and partially compensate one another, which is precisely the structure expected near marginality, where the global growth rate is the small residual of competing local contributions. Meanwhile, the imaginary parts (dashed curves) remain comparatively confined to the upstream region for all three cases, indicating that the oscillation frequency is still set predominantly by the near-inlet feedback core even though viscoelasticity distributes the sensitivity farther downstream. Taken together, figure~\ref{fig:fig3} supports two conclusions. First, all three configurations behave as genuine global oscillators, i.e., both wavemaker and endogeneity decay well before the outlet, indicating that the selected mode is not an artefact of domain truncation. Second, viscoelasticity modifies the instability mechanism in a specific and important way: as $\De$ increases, the feedback-sensitive region becomes progressively more widespread while a distinct polymeric endogeneity contribution appears, showing that the marginal state is controlled by a spatially distributed capillary--elastic feedback loop rather than by a purely Newtonian, near-nozzle mechanism.

\section{Influence of rheology on marginal conditions}
\label{sec:marginal}

\begin{figure}
    \centering
    \includegraphics[width=1.00\linewidth]{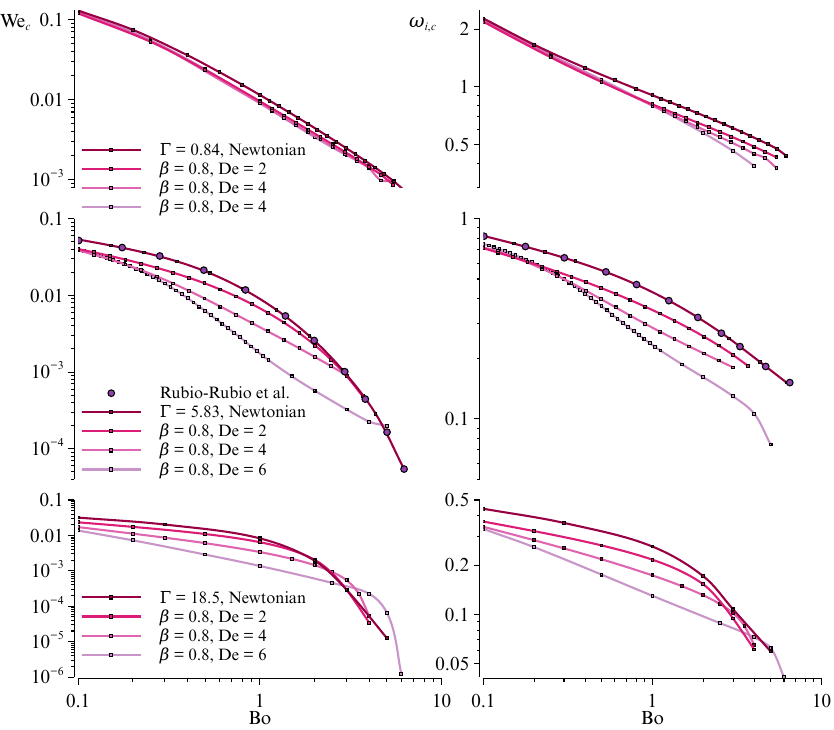}
    \caption{Marginal curves $\We_c(\Bo)$ and corresponding frequencies $\omega_{i,c}(\Bo)$ for $\Gamma = 0.84, 5.83$, and $18.5$.}
    \label{fig:fig4}
\end{figure}

Figure~\ref{fig:fig4} generalizes the marginal-curve analysis by varying the viscous level over three representative Kapitza numbers, $\Gamma=0.84$, $5.83$ and $18.5$, while keeping the viscosity ratio fixed at $\beta=0.8$ and sweeping $\De$. Each row reports the critical Weber number $\We_c(\Bo)$ at the onset of the self-excited global oscillation and the corresponding critical frequency $\omega_{i,c}(\Bo)$. For all three values of $\Gamma$, the Newtonian reference curve decreases monotonically with $\Bo$ in both panels. Increasing the nozzle size, and therefore strengthening gravity relative to surface tension, shifts the transition to progressively smaller flow rates and to lower oscillation frequencies. This indicates that the global feedback loop becomes slower as the jet is more strongly stretched downstream. The middle row, $\Gamma=5.83$, includes the Newtonian data of \citet{RubioRubio2013} and recovers both the marginal branch and the onset frequency quantitatively, thereby extending the Newtonian benchmark of \S4.1 to the full marginal curve. 

Superimposed on this Newtonian baseline, viscoelasticity produces a systematic ordering with $\De$ across the whole $(\Bo,\Gamma)$ range considered. For fixed $(\Bo,\Gamma)$, increasing $\De$ shifts both $\We_c$ and $\omega_{i,c}$ downward, so that $\We_c(\De=2)>\We_c(\De=4)>\We_c(\De=6)$ and $\omega_{i,c}(\De=2)>\omega_{i,c}(\De=4)>\omega_{i,c}(\De=6)$, with the viscoelastic curves lying below the Newtonian one. In the present formulation this trend is interpreted in terms of an additional tensile-stress pathway. As $\De$ increases at fixed $\beta$, polymeric stresses are transported over longer axial distances before relaxing. This strengthens the base-state tensile resistance and enhances the linear coupling by which stress perturbations feed back on the free surface through the axial-tension term. In effect, elastic tension competes with capillary-driven thinning and delays the onset of the global oscillatory instability, so that marginality is reached only when the through-flow is reduced, meaning smaller $\We$. The reduction in $\omega_{i,c}$ with $\De$ is consistent with the same picture. Marginality occurs at weaker advection and the polymer introduces a memory time scale that adds phase lag to the global loop. Both effects promote slower oscillations at onset. This interpretation is consistent with the wavemaker and endogeneity diagnostics of \S\ref{sec:wavemaker}, which show that elasticity broadens the feedback-sensitive region and introduces a non-negligible polymeric contribution to eigenvalue production in marginal configurations.

Varying $\Gamma$ modulates these trends in a physically informative way. Increasing $\Gamma$ shifts the overall Newtonian boundary downward in $\We_c$ and in $\omega_{i,c}$, which reflects the stabilizing role of viscous stresses in suppressing capillary-driven unsteadiness and permitting sustained jetting at lower flow rates. The incremental viscoelastic shift is noticeably weaker for the lowest viscous level, $\Gamma=0.84$, where the curves nearly collapse and the dependence on $\De$ becomes mild. This near-collapse is consistent with a regime in which onset is controlled by a rapid inertio-capillary balance localized near the inlet, leaving limited opportunity for polymeric stresses to accumulate and influence the marginal condition. It is also the regime where \citet{RubioRubio2013} emphasize that additional upstream physics can become important for low-viscosity liquids issued through long injectors, in particular viscous relaxation of the exit velocity profile \citep[see also][]{sevilla2011effect}. By way of contrast, for $\Gamma=5.83$ and $18.5$ the viscoelastic shift is clearly resolved, especially at small-to-moderate $\Bo$, and it weakens as $\Bo$ increases. This is consistent with the increasing dominance of gravity-driven stretching and advection at large $\Bo$, which drives the marginal curves back towards the Newtonian scaling. Overall, figure~\ref{fig:fig4} shows that moderate elasticity, $\De=O(1)$, induces systematic shifts of both the critical flow rate and the onset frequency, and that these shifts depend on viscosity in a manner consistent with a coupled capillary--elastic feedback loop whose influence is strongest when the jet is not too strongly gravity dominated.

\section{Conclusions}
\label{sec:conclusions}

We have formulated and analysed a one-dimensional model for gravitationally stretched viscoelastic jets that combines a full-curvature slender-jet description with a Giesekus stress closure. The framework retains the spatial development of the base state and admits a global eigenvalue formulation. This provides a unified route to predict the onset of the jetting--dripping transition for weakly elastic liquids, and to interpret the instability mechanism using direct--adjoint diagnostics that quantify receptivity, structural sensitivity and endogeneity.

In the Newtonian limit we reproduced the benchmark results of \citet{RubioRubio2013}, including marginal spectra, base-state profiles, and the dependence of the critical flow rate and onset frequency on Bond and Kapitza numbers. This validation supports the accuracy of the numerical implementation and confirms that the onset is a genuine global instability selected by the inlet and the non-parallel base flow rather than a truncation artefact. Extending this baseline to viscoelastic liquids, we identified robust shifts of the leading Hopf eigenpair as elasticity is increased. For fixed $(\Bo,\Gamma)$ and viscosity ratio $\beta$, increasing the Deborah number decreases both the critical Weber number and the selected onset frequency. This trend persists across the three viscous levels considered, and it is most pronounced at small-to-moderate $\Bo$ while weakening as $\Bo$ increases and gravity-driven stretching becomes dominant.

The direct--adjoint diagnostics provide a mechanistic interpretation of these parametric trends. In Newtonian jets the wavemaker is sharply localized near the inlet and the endogeneity is dominated by the continuity and capillarity pathway, with momentum contributions remaining comparatively small in marginal configurations. When viscoelasticity is introduced, the instability retains its upstream-selected global-oscillator character, but the feedback-sensitive region spreads downstream in a systematic way as $\De$ increases. This spreading reflects the advection and relaxation of polymeric stresses, and it is accompanied by a distinct polymeric contribution to eigenvalue production. In marginal cases, the real parts of the continuity-driven and stress-driven endogeneities partially compensate, which indicates that onset is governed by a coupled capillary--elastic feedback loop rather than by a purely Newtonian balance. The imaginary components remain concentrated closer to the inlet, which supports the view that the oscillation frequency is set predominantly by the near-nozzle core even when viscoelasticity distributes sensitivity farther downstream.

Several extensions follow naturally. First, the inlet-stress closure can be linked to simple die models or to measured swell ratios, and its influence can be quantified systematically using the same receptivity tools developed here. Second, the present framework is well suited to resolvent and forced-response analyses in order to connect global stability with noise amplification and experimentally observed intermittency near marginality. Third, incorporating additional physics such as viscous relaxation of the exit velocity profile, weak confinement, or surface-active effects would allow a more direct comparison with experiments in low-viscosity regimes and could clarify the interplay between axial curvature stabilization and viscoelastic tension. More broadly, the results suggest that one of the most consequential dynamical effects of weak elasticity in stretched jets is not solely an eigenvalue drift, but a redistribution of the feedback-sensitive region along the jet, which has direct implications for modelling, reduced-order descriptions, and control strategies for jetting and dispensing processes.



\backsection[Acknowledgements]{Professors M. Rubio-Rubio and A. Sevilla are thanked for fruitful discussions.}

\backsection[Funding]{This work has been supported by the Spanish MINECO under project PID2020-115655GB-C22, partly financed through FEDER European funds.}

\backsection[Declaration of interests]{The authors report no conflict of interest.}


\backsection[Author ORCIDs]{D. Moreno-Boza, https://orcid.org/0000-0002-8663-0382}



\appendix

\section{Explicit linearized equations }\label{appA}

We linearize the dimensionless governing equations \eqref{eq:mass_nondim}--\eqref{eq:taurr_nondim} about a steady base state
$\bm{q}_0(z)=[r_0,u_0,\tau_{zz,0},\tau_{rr,0}]^{\mathrm{T}}$ and seek normal modes
$\hat{\bm{q}}(z)=[\hat r,\hat u,\hat\tau_{zz},\hat\tau_{rr}]^{T}$ such that
$\bm{q}(z,t)=\bm{q}_0(z)+\epsilon\,\hat{\bm{q}}(z)\mathrm{e}^{\omega t}$, with $\epsilon\ll 1$.
Throughout, primes denote derivatives with respect to $z$. When the perturbation enters through a nonlinear functional of the base state (e.g.\ curvature or conservative fluxes), we employ the first variation (Fr\'echet derivative) notation
\[
\delta F \equiv \left.\frac{\mathrm{d}}{\mathrm{d}\epsilon}\,F\!\left[r_0+\epsilon\hat r,\;u_0+\epsilon\hat u,\;\tau_{zz,0}+\epsilon\hat\tau_{zz},\;\tau_{rr,0}+\epsilon\hat\tau_{rr}\right]\right|_{\epsilon=0}.
\]
With this convention, the linearized system can be written compactly as $(\bm{A}-\omega\bm{B})\hat{\bm{q}}=\bm{0}$, where $\bm{B}$ collects the temporal (mass-matrix) terms and $\bm{A}$ collects spatial derivatives and couplings. Thus, the linearised mass conservation and momentum laws read
\begin{equation}
\omega\,(2r_0\hat r)+\partial_z\!\left(2r_0u_0\hat r+r_0^{2}\hat u\right)=0,
\label{eq:lin_mass_app}
\end{equation}
\begin{equation}
\omega\hat u+u_0\hat u' + u_0'\hat u
= -\partial_z(\delta C)
+\Gamma\,\delta\!\left[\frac{1}{r^2}\partial_z\!\left(r^2 u'\right)\right]
+\frac{\Gamma(1-\beta)}{3\beta}\,
\delta\!\left[\frac{1}{r^2}\partial_z\!\left(r^2 T\right)\right],
\label{eq:lin_mom_app}
\end{equation}
where $T=\tau_{zz}-\tau_{rr}$ and $\delta C$ is the first variation of the full curvature \eqref{eq:curvature}. The explicit expression for $\delta C$ follows by differentiating the curvature functional~\eqref{eq:curvature}. Introducing
$s_0=(1+r_0'^2)^{-1/2}$, we obtain
\begin{equation}
\delta C
= -\frac{s_0}{r_0^{2}}\hat r
-\frac{r_0's_0^{3}}{r_0}\hat r'
-s_0^{3}\hat r''
+3\,r_0''\,r_0'\,s_0^{5}\,\hat r'.
\label{eq:dC_app}
\end{equation}
For implementation purposes, it is also useful to expand the conservative flux variations appearing in \eqref{eq:lin_mom_app}. Denoting $\hat T=\hat\tau_{zz}-\hat\tau_{rr}$, one convenient form is
\begin{align}
\delta\!\left[\frac{1}{r^2}\partial_z\!\left(r^2 u'\right)\right]
&=
-\frac{2\hat r}{r_0^{3}}\partial_z\!\left(r_0^{2}u_0'\right)
+\frac{1}{r_0^{2}}\partial_z\!\left(2r_0\hat r\,u_0' + r_0^{2}\hat u'\right),
\label{eq:dvisc_app}\\[4pt]
\delta\!\left[\frac{1}{r^2}\partial_z\!\left(r^2 T\right)\right]
&=
-\frac{2\hat r}{r_0^{3}}\partial_z\!\left(r_0^{2}T_0\right)
+\frac{1}{r_0^{2}}\partial_z\!\left(2r_0\hat r\,T_0 + r_0^{2}\hat T\right),
\label{eq:dtens_app}
\end{align}
with $T_0=\tau_{zz,0}-\tau_{rr,0}$. 
Finally, the diagonal Giesekus constitutive laws~\eqref{eq:tauzz}--\eqref{eq:taurr} linearise to
\begin{align}
\De\Big(\omega\hat\tau_{zz} + u_0\hat\tau_{zz}' + \hat u\,\tau_{zz,0}'
-2u_0'\hat\tau_{zz}-2\hat u'\tau_{zz,0}\Big)
+\hat\tau_{zz}+2\alpha\De\,\tau_{zz,0}\hat\tau_{zz}
&=2\hat u',
\label{eq:lin_tzz_app}\\[4pt]
\De\Big(\omega\hat\tau_{rr} + u_0\hat\tau_{rr}' + \hat u\,\tau_{rr,0}'
+u_0'\hat\tau_{rr}+\hat u'\tau_{rr,0}\Big)
+\hat\tau_{rr}+2\alpha\De\,\tau_{rr,0}\hat\tau_{rr}
&=-\hat u'.
\label{eq:lin_trr_app}
\end{align}
In the Newtonian limit $\De=0$ (or $1-\beta=0$ in the momentum coupling), these constitutive equations are dropped and the tensile-stress contribution in \eqref{eq:lin_mom_app} vanishes.

The eigenmode boundary conditions follow from the fact that inlet data are prescribed; thus, perturbations satisfy homogeneous inlet conditions,
\begin{equation}
\hat r(0)=0,\qquad \hat u(0)=0,\qquad \hat\tau_{zz}(0)=0,\qquad \hat\tau_{rr}(0)=0.
\label{eq:lin_bc_inlet_app}
\end{equation}
At the downstream boundary $z=L$, we impose homogeneous outflow conditions consistent with the steady integration (e.g.\ the homogeneous counterpart of the ``straight jet'' condition $\hat{r}_z=\hat{r}_{zz}=0$). For sufficiently large $L$, the leading eigenvalues are insensitive to the specific admissible outflow choice provided the instability core remains well upstream of the truncation.


\bibliographystyle{jfm}
\bibliography{jfm}

\end{document}